\documentclass[preprint,10pt,a4paper,final,tightenlines,showpacs,bibnotes,nofootinbib,notitlepage,pra,twocolumn]{revtex4}
\usepackage{amssymb}

\usepackage{amsmath}

\newtheorem{theorem}{Theorem}
\newtheorem{acknowledgement}[theorem]{Acknowledgement}

\input{tcilatex}

\begin{document}

\title{Initial Decoherence and Disentanglement of Open Two-Qubit Systems}
\author{Xian-Ting Liang\thanks{%
Electronic address: xtliang@ustc.edu}}
\affiliation{Department of Physics and Institute of Modern Physics, Ningbo University,
Ningbo, 315211, China}

\begin{abstract}
In this paper we investigate a open two-qubit model whose dynamics is not
exactly solvable. When the initial state is the maximum entangled state, as
the exactly solvable open two-qubit model [D. Tolkunov and V. Privman, Phys.
Rev. A 71, 060308(R) (2005)], the decay of entanglement of formation of the
model, expressed by concurrence is also governed by the product of
suppression factors describing decoherence of the subsystems (qubits).
However, if the initial state is not the maximum entangled state, its
concurrence will decrease faster than the product of the suppression factors
describing decoherence of the qubits.

\end{abstract}

\maketitle

\section{Introduction}

Coherence and entanglement are two remarkable features of quantum systems,
and they are also dominating reasons why quantum computation and quantum
communication differ from the classical ones \cite{Nielsen-Chuang}. For
example, quantum coherence can lead to natural parallel computations which
can enhance efficiency for solving some complex problems by using effective
quantum algorithms. Quantum bit (qubit) is a key block for building quantum
computers. However, the interactions of qubits with their environment will
induce loss of the coherence, decoherence. Decoherence is considered a
central impediment to fabricate quantum computers. In quantum computation
and quantum communication, the entanglement, nonlocal correlation between
the quantum subsystems is also needed. The entanglement is recognized as an
important physical resource in quantum information transmission and
processing. Many protocols in quantum communication and quantum computation
are based on the entangled states. It is also shown that couplings of the
quantum systems and its subsystems to their environment will result in the
loss of the entanglement, and this loss cannot be restored by local
operations and classical communications \cite{Bennett,Peres}. Both the
decoherence and the loss of the entanglement may result from the
interactions of the quantum systems with their environment. Then, what
relations are there between the two quantities? In this paper we will
investigate the relationship of the decoherence and the loss of entanglement
for a non-exactly solvable two-qubit model. As Ref.\cite%
{Privman-Tolkunov-1,Privman-Tolkunov-2}, let us refer to two subsystems, $%
S^{\left( 1\right) }$ and $S^{\left( 2\right) },$ of the combined system, $%
S. $ The evolutions of the coherence of the subsystems $S^{\left( 1\right) }$
and $S^{\left( 2\right) }$ in their bathes can be described by suppression
factors, $0\leq \delta ^{\left( 1,2\right) }\left( t\right) \leq 1.$ On the
other hand, the evolutions of the entanglement between the subsystems $%
S^{\left( 1\right) }$ and $S^{\left( 2\right) }$ can be described by
entanglement of formation \cite{Bennett02} expressed by concurrence \cite%
{Concurrence}. In large times the decay of the coherent terms are
exponential, then the decay rate of the whole system is the summation of the
rates of the subsystems \cite{Storcz-Wilhelm,Ischi-Hilke-Dube,cpsun}$.$ Yu
and Eberly \cite{Yuetal01} found that for a exactly solvable model in which
the decoherence is caused by pure dephasing process, the concurrence decays
faster than the quantum dephasing of an individual qubit. A more physical
model, two entangled atoms in pure vacuum noise, is investigated recently by
the same authors \cite{Yuetal02}. They found that the disentanglement time
is shorter than the usual spontaneous lifetime. V. Privman \emph{et al.} %
\cite{Privman-Tolkunov-1,Privman-Tolkunov-2} investigated another
dynamically solvable model. They shown that the \emph{decay of concurrence}
of the model is governed by the product of suppression factors describing
decoherence of the subsystems (qubits) in a short time. It is interesting
that whether or not the relationship of the concurrence and the suppression
factors describing the decoherence can be held for other open quantum
systems. In this paper we shall investigate a different open two-qubit model
whose dynamics is not exactly solvable. By using a short-time approximation
we can obtain the evolution of reduced density matrix of the system. It will
be shown that as the initial state of the two-qubit system is the maximum
entangled states the decay of concurrence, namely the decay of the degree of
entanglement of formation can also be governed by the product of suppression
factors describing decoherence of the subsystems (qubits) in a short time.
If the initial state is not the maximum entangled state, the concurrence
will decrease faster than the product of the suppression factors describing
the coherence of the qubits does.

\section{Decoherence and the loss of entanglement}

Suppose the open two-qubit system has Hamiltonian%
\begin{equation}
H=\sum_{r=1}^{2}\left( H_{s}^{r}+H_{B}^{r}+H_{I}^{r}\right) ,  \label{e1}
\end{equation}%
where%
\begin{eqnarray}
H_{s}^{r} &=&-\frac{1}{2}E_{J}^{r}\sigma _{x}^{r},  \notag \\
H_{B}^{r} &=&\sum_{k}M_{k}^{r}=\sum_{k}\omega _{k}^{r}b_{k}^{r\dagger
}b_{k}^{r},  \notag \\
H_{I}^{r} &=&\sigma _{z}^{r}\sum_{k}\left( g_{k}^{r\ast
}b_{k}^{r}+g_{k}^{r}b_{k}^{r\dagger }\right) .  \label{e2}
\end{eqnarray}%
Here, we use the subscripts or the superscripts $r=1,2$ to label the qubits. 
$H_{s}^{r}$ and $H_{B}^{r}$ are the Hamiltonian of qubits and their bosonic
bathes and $H_{I}^{r}$ are the interactions of the qubits with their bathes %
\cite{Leggett}. Where we assume that each qubit interacts with its own bath.
This is not a exactly solvable model. If we do not consider the interactions
between the qubits the evolution operator of the combined system can be
expressed as%
\begin{equation}
U=U_{1}\otimes U_{2}.  \label{e3}
\end{equation}%
The evolution operator of the single qubit is%
\begin{equation}
U_{r}=e^{-iH^{r}\tau /\hslash }=e^{-i\left(
H_{s}^{r}+H_{I}^{r}+H_{B}^{r}\right) t},  \label{e4}
\end{equation}%
where $t=\tau /\hslash .$ Due to non-conservation of $H_{s}$ in this system,
the evolution operator cannot be in a general way expressed as $%
e^{-iH_{s}^{r}t}e^{-i\left( H_{I}^{r}+H_{B}^{r}\right) t}.$ But in the
sort-time approximation, the operator can be approximately expressed as \cite%
{split-operator01,split-operator02}%
\begin{equation}
U_{r}=e^{-iH_{s}^{r}t/2}e^{-i\left( H_{I}^{r}+H_{B}^{r}\right)
t}e^{-iH_{s}^{r}t/2}+o(t^{3}).  \label{e5}
\end{equation}%
It has been proved that the expression is accurate enough as the time being
short to the characteristic time \cite{Privman}. So the elements of the
reduced density matrix $\rho \left( t\right) $ in the basis of operator $%
H_{s}$ can be expressed as%
\begin{eqnarray}
\rho _{mn}^{r} &=&\text{Tr}_{B}\left\langle \varphi _{m}\right|
e^{-iH_{s}^{r}t/2}e^{-i\left( H_{I}^{r}+H_{B}^{r}\right)
t}e^{-iH_{s}^{r}t/2}R\left( 0\right)  \notag \\
&&e^{iH_{s}^{r}t/2}e^{i\left( H_{I}^{r}+H_{B}^{r}\right)
t}e^{iH_{s}^{r}t/2}\left| \varphi _{n}\right\rangle .  \label{e6}
\end{eqnarray}%
Here, we suppose the initial state of the system be $R^{r}\left( 0\right)
=\rho ^{r}\left( 0\right) \otimes \Theta ^{r}$ where $\rho ^{r}\left(
0\right) $ and $\Theta ^{r}$ are initial states of the qubit and its
environment. The latter is the product of the bath modes density matrices\ $%
\theta _{k}^{r}$. In the initial states, each bath mode $k$ is assumed to be
thermalized, namely,%
\begin{equation}
\theta _{k}^{r}=\frac{e^{-\beta M_{k}^{r}}}{\text{Tr}_{k}\left( e^{-\beta
M_{k}^{r}}\right) },  \label{e7}
\end{equation}%
where $\beta =1/kT$, $T$ is the temperature and $k$ is the Boltzmann
constant.

\subsection{Decoherence and decay of coherence}

In this subsection we shall investigate the decoherence of each qubit due to
the interaction of the qubit with its own environment, a bath. Here, we
denote the suppression factors describing the decoherence of each qubits
with $\delta ^{r}$. The relationship of the suppression factors $\delta ^{r}$
with the usual term decoherence $D^{r}$ is $D^{r}=L\left( 1-\delta
^{r}\right) ,$ where $L$ is the initial coherent terms of the density
matrix, namely, the off-diagonal elements of the density matrix of the
initial state for the $r-th$ qubit. $L=1/2$ for the initial states with
maximum coherent terms. We call $L\delta ^{r}$ the \emph{decay of coherence}
of the $r-th$ qubit. Usually, the environment is assumed to be a large
macroscopic the interaction with it leads to the thermal equilibrium at
temperature $T$. In this case, Markovian type approximations can be used to
quantified the decoherent process and it usually yields the exponential
decay of the density matrix elements in the energy basis of the Hamiltonian $%
H_{s}^{r}.$ Thus, the decay rates will be additive. In this time scale the
measures of entropy and the first entropy can be used for quantifying the
decoherence. But the decoherence of the qubit gate operations cannot be
characterized by this methods because the time of the elementary quantum
gate operation is much shorter than the thermal relaxation time. It has been
shown that the norms $\left\| \sigma ^{r}\right\| _{\lambda }$ is useful for
describing the decoherence of the short-time evolution \cite{Privman}. Here $%
\sigma ^{r}$ is the deviation operator defined as 
\begin{equation}
\sigma ^{r}\left( t\right) =\rho ^{r}\left( t\right) -\rho _{i}^{r}\left(
t\right) ,  \label{e8}
\end{equation}%
where $\rho ^{r}\left( t\right) $ and $\rho _{i}^{r}\left( t\right) $ are
density matrixes of the ``real''\ evolution (with interaction) and the
``ideal''\ one (without interaction) of the $r-th$ qubit. We can use the
norm $\left\| \sigma ^{r}\right\| _{\lambda }$ describing the decoherence
and the norm is defined as \cite{Privman}%
\begin{equation}
\left\| \sigma ^{r}\right\| _{\lambda }=\sup_{\varphi \neq 0}\left( \frac{%
\left\langle \varphi \right| \sigma ^{r}\left| \varphi \right\rangle }{%
\left\langle \varphi \right. \left| \varphi \right\rangle }\right) ^{\frac{1%
}{2}}.  \label{e9}
\end{equation}%
For a qubit, the norm can be given by 
\begin{equation}
\left\| \sigma ^{r}\right\| _{\lambda }=\sqrt{\left| \sigma _{10}^{r}\right|
^{2}+\left| \sigma _{11}^{r}\right| ^{2}}.  \label{e10}
\end{equation}%
It is shown that for a given system, the norm $\left\| \sigma ^{r}\right\|
_{\lambda }$ increase with time, reflecting the decoherence of the system.
However, in general it is oscillated at the system's internal frequency.
Thus, the decohering effect of the bath is better quantified by the maximal
operator norm 
\begin{equation}
D^{r}\left( t\right) =\sup_{\rho ^{r}\left( 0\right) }\left( \left\| \sigma
^{r}\left( t,\rho ^{r}\left( 0\right) \right) \right\| _{\lambda }\right) .
\label{e11}
\end{equation}%
For our investigating model, by using Eqs.(\ref{e6}) and \ref{e11}) we can
obtain the decoherence of the qubit as 
\begin{equation}
D^{r}\left( t\right) =\frac{1}{2}\left( 1-e^{-4G^{r}\left( t\right) }\right)
,  \label{e12}
\end{equation}%
where 
\begin{equation}
G^{r}\left( t\right) =2\tsum_{k}\frac{\left| g_{k}^{r}\right| ^{2}}{\omega
_{k}^{r2}}\sin ^{2}\frac{\omega _{k}^{r}t}{2}\coth \frac{\beta \omega
_{k}^{r}}{2}.  \label{e13}
\end{equation}%
It is shown that the suppression factor describing the decoherence of the
open qubit $r$ is%
\begin{equation}
\delta ^{r}=e^{-4G^{r}\left( t\right) }.  \label{eq14}
\end{equation}

\subsection{Loss of the entanglement and decay of the entanglement}

Numerous measures of entanglement have been considered over the years. For
quantum information content, the entanglement of formation has been a widely
accepted measure of entanglement. The measure can measure the degree of
entanglement not only for pure states but also for mixed states. The
concurrence is a quantity monotonically related to the entanglement of
formation. For a pure or mixed state, $\rho _{s},$ of two qubits, one can
define the spin-flipped state,%
\begin{equation}
\tilde{\rho}_{s}=\left( \sigma _{y}\otimes \sigma _{y}\right) \rho
_{s}^{\ast }\left( \sigma _{y}\otimes \sigma _{y}\right) ,  \label{e15}
\end{equation}%
and the Hermitian matrix,%
\begin{equation}
R\left( \rho _{s}\right) =\sqrt{\sqrt{\rho _{s}}\tilde{\rho}_{s}\sqrt{\rho
_{s}}},  \label{e16}
\end{equation}%
with eigenvalues $\lambda _{i=1,2,3,4}$. Here, $\rho _{s}^{\ast }$ denotes
the complex conjugation of $\rho $ in the standard basis and $\sigma _{y}$
is one of the Pauli matrixes. The concurrence, $C$, is defined by \cite%
{Concurrence}%
\begin{equation}
C=\max \left\{ 0,2\max_{i}\lambda _{i}-\sum_{j=1}^{4}\lambda _{j}\right\} .
\label{e17}
\end{equation}%
From Eq.(\ref{e6}) and helped with operator-algebra techniques we can obtain
the evolutions of the reduced density matrix as%
\begin{equation}
\rho ^{r}\left( t\right) =\left( 
\begin{tabular}{ll}
$\rho _{00}^{r}\left( t\right) $ & $\rho _{01}^{r}\left( t\right) $ \\ 
$\rho _{10}^{r}\left( t\right) $ & $\rho _{11}^{r}\left( t\right) $%
\end{tabular}%
\right) ,  \label{e18}
\end{equation}%
where%
\begin{eqnarray}
\rho _{00}^{r}\left( t\right) &=&\frac{1}{2}\rho _{00}^{r}\left(
1+e^{-4G^{r}\left( t\right) }\right) +\frac{1}{2}\rho _{11}^{r}\left(
1-e^{-4G^{r}\left( t\right) }\right) ,  \notag \\
\rho _{01}^{r}\left( t\right) &=&\frac{e^{-itE_{J}^{r}}}{2}\rho
_{01}^{r}\left( 1+e^{-4G^{r}\left( t\right) }\right) +\frac{1}{2}\rho
_{10}^{r}\left( 1-e^{-4G^{r}\left( t\right) }\right) ,  \notag \\
\rho _{01}^{r}\left( t\right) &=&\frac{1}{2}\rho _{01}^{r}\left(
1-e^{-4G^{r}\left( t\right) }\right) +\frac{e^{-itE_{J}^{r}}}{2}\rho
_{10}^{r}\left( 1+e^{-4G^{r}\left( t\right) }\right) ,  \notag \\
\rho _{11}^{r}\left( t\right) &=&\frac{1}{2}\rho _{00}^{r}\left(
1-e^{-4G^{r}\left( t\right) }\right) +\frac{1}{2}\rho _{11}^{r}\left(
1+e^{-4G^{r}\left( t\right) }\right) .  \label{e19}
\end{eqnarray}%
In the right hand side of Eq.(\ref{e19}) we denote $\rho _{ij}^{r}\left(
0\right) $ with $\rho _{ij}^{r}.$ If we do not consider the interactions
between the qubits, the reduced density matrix of the combined system $S$
becomes%
\begin{equation}
\rho _{s}\left( t\right) =\rho ^{1}\left( t\right) \otimes \rho ^{2}\left(
t\right) .  \label{e20}
\end{equation}%
In the following, we set the system in a pure entangled state%
\begin{equation}
\rho _{s}\left( 0\right) =\left| \Psi \right\rangle \left\langle \Psi \right|
\label{e21}
\end{equation}%
at $t=0$, where%
\begin{equation}
\left| \Psi \right\rangle =\frac{1}{\sqrt{1+\left| \alpha \right| ^{2}}}%
\left[ \left| 01\right\rangle +\alpha \left| 10\right\rangle \right] .
\label{e22}
\end{equation}%
The concurrence of the initial state is $C(0)=2\left| \alpha \right| /\left(
1+\left| \alpha \right| ^{2}\right) .$ At first, we investigate an especial
case, namely, the case of $\alpha =1$, which denotes the initial state is a
maximum entangled state. We can easily obtain that after time $t$ the
density matrix of the open two-qubit system becomes%
\begin{equation}
\rho _{s}\left( t\right) =\frac{1}{4}\left[ 
\begin{tabular}{llll}
$A$ & $0$ & $0$ & $Ae^{-itE_{J}}$ \\ 
$0$ & $B$ & $B$ & $0$ \\ 
$0$ & $B$ & $B$ & $0$ \\ 
$Ae^{itE_{J}}$ & $0$ & $0$ & $A$%
\end{tabular}%
\right] .  \label{e23}
\end{equation}%
where $A=1-e^{-4G^{1}\left( t\right) -4G^{2}\left( t\right) }$, and $%
B=1+e^{-4G^{1}\left( t\right) -4G^{2}\left( t\right) }.$ Then, we can obtain
the eigenvalues of the product $\rho _{s}\left( t\right) \tilde{\rho}%
_{s}\left( t\right) $ as%
\begin{equation}
\mu _{1,2}=\frac{1}{4}\left( 1\pm 2e^{-4G^{1}\left( t\right) -4G^{2}\left(
t\right) }+e^{-8G^{1}\left( t\right) -8G^{2}\left( t\right) }\right) ,
\label{e24}
\end{equation}%
and%
\begin{equation}
\mu _{3,4}=0.  \label{e25}
\end{equation}%
Finally, we can obtain the concurrence as%
\begin{equation}
C\left( t\right) =\left| \sqrt{\mu _{1}}-\sqrt{\mu _{2}}\right|
=e^{-4G^{1}\left( t\right) -4G^{2}\left( t\right) }=C(0)\delta ^{1}\left(
t\right) \delta ^{2}\left( t\right) .  \label{e26}
\end{equation}%
We call the $C\left( t\right) $ the decay of entanglement. The loss of the
entanglement is $1-C(t)$. If we set $\alpha =-1$ or $\pm i$ we can obtain
some similar results. It is very interesting that in this case the
suppression factors describing the decoherence of the qubits and the
entanglement of formation between the two qubits still has the compact
relationship as Ref.\cite{Privman-Tolkunov-1,Privman-Tolkunov-2}. But, if
the relationship is still preserved for other initial states? In the
following we shall investigate the cases that the initial states are not the
maximum entangled states. When $\alpha \neq \pm 1$ or $\pm i$ the operator
calculations similar to above derivation becomes very complex, then we only
be able to numerically investigate this problem with a concrete bath model.
Suppose the two bathes with Ohmic noise spectrum are the same for
convenience, which do not loss the universality. The spectral densities of
the bathes may be expressed as%
\begin{equation*}
J\left( \omega \right) =\eta \omega e^{-\omega /\omega _{c}}.
\end{equation*}%
It is well-known that when the summation in Eq.(\ref{e13}) is converted to
the integration in the limit of infinite number of the bath modes, one has%
\begin{equation*}
G^{r}\left( t\right) =2\eta \int d\omega e^{-\omega /\omega _{c}}\omega
^{-1}\sin ^{2}\frac{\omega t}{2}\coth \frac{\beta \omega }{2}
\end{equation*}%
for the real $g\left( \omega \right) .$ 
\begin{eqnarray*}
&& \\
&& \\
&&Fig.1 \\
&& \\
&&
\end{eqnarray*}%
In this paper we choose the cutoff frequency of the bath modes as $\omega
_{c}=10^{12}$ $Hz.$ It is pointed that the dimensionless strength of the
dissipation is very small for some qubit environment. For example, for the
Josephson charge qubit in Ohmic bath the value of $\eta $ is about $10^{-6}$ %
\cite{chemphys,chemphys295}$.$ In our calculations, if we take $\eta
=10^{-6},$ the difference of the decay of concurrence $C\left( t\right) $
and the\ product of the suppression factors $\delta ^{1}\left( t\right)
\delta ^{2}\left( t\right) $ with a initial concurrence $C(0)$ is very
small. For convenience, where we denote $S(t):=C(0)\delta ^{1}\left(
t\right) \delta ^{2}\left( t\right) .$ When we increase $\eta $ to $10^{-5}$
we can clearly see that the decay of concurrence decreases faster than $S(t)$
does. In Fig.1 we plot $C\left( t\right) $ versus time (with points), which
is compared with $S(t)$ (with lines) in different initial states where (a) $%
\alpha =1,$ (b) $\alpha =2,$ and (c) $\alpha =3$. It is shown that the
concurrence is not equals to the product of suppression factors and its
initial concurrence, namely $S(t)$ except for $\alpha =1$. In plotting of
the figure we choose a time unit, $ks$ for convenience. Where we set $1$ $%
ks=1519.29$ $ps$. From Fig.1 we see that the durative time of our
calculations is $\tau =12.15ps$, which is smaller than the characteristic
time of the qubits \cite{Liang}. So our calculations is accurate enough.

It is shown that system%
\begin{equation*}
\tilde{H}=\sum_{r=1}^{2}\left( \tilde{H}_{s}^{r}+H_{B}^{r}+\tilde{H}%
_{I}^{r}\right) ,
\end{equation*}%
where%
\begin{eqnarray}
\tilde{H}_{s}^{r} &=&-\frac{1}{2}E_{J}^{r}\sigma
_{z}^{r},H_{B}^{r}=\sum_{k}\omega _{k}^{r}b_{k}^{r\dagger }b_{k}^{r},  \notag
\\
\tilde{H}_{I}^{r} &=&\sigma _{x}^{r}\sum_{k}\left( g_{k}^{r\ast
}b_{k}^{r}+g_{k}^{r}b_{k}^{r\dagger }\right) ,  \label{e28}
\end{eqnarray}%
is also a non-exactly solvable model and it has the same dynamics to system
Eqs.(\ref{e1}-\ref{e2}) in short-time approximation \cite{Privman,Liang}. So
the concurrence and the suppression factors describing the decoherence of
the system may have a similar relationship to system Eqs.(\ref{e1}-\ref{e2}).

\section{Conclusions}

In this paper we investigated a non-exactly solvable open two-qubit model.
We obtained that in a short time the decay of entanglement of formation is
governed by the product of the suppression factors describing decoherence of
the subsystems as the initial state of the two-qubit system is the maximum
entangled state. This novel relationship is similar to the discover by V.
Privman \emph{et al.} in \cite{Privman-Tolkunov-1,Privman-Tolkunov-2}, where
the open two-qubit system is exactly solvable. Our work shows that when the
initial state is not a maximum entangled state, after the open two-qubit
evolve a short time $t$ the entanglement of formation, concurrence decrease
faster than the product of the suppression factors describing the
decoherence of the two-qubit system, which is similar to the discover by Yu
and Eberly in \cite{Yuetal01,Yuetal02}. It is shown that when the
dissipation is not very weak (the dimensionless strength of the dissipation $%
\eta \gtrapprox 10^{-6}$) the entanglement is distinctly more frangible than
the coherence for the quantum systems.

\begin{acknowledgement}
The project was supported by National Natural Science Foundation of China
(Grant No. 10347133) and Ningbo Youth Foundation (Grant No. 2004A620003).
\end{acknowledgement}

\section{Caption of Fig.1}

Fig.1: The concurrence $C(\alpha ,t)$ (points) and $S(\alpha ,t)=2\alpha
/(1+\left| \alpha \right| ^{2})\delta ^{1}\left( t\right) \delta ^{2}\left(
t\right) $ (lines) of a two-qubit system in their Ohmic bathes for different
initial states (a) $\alpha =1,$ (b) $\alpha =2,$ (c) $\alpha =3$. Here we
take dimensionless strength of the dissipation $\eta =1\times 10^{-5},$ the
cutoff frequencies of the bath modes $\omega _{c}=10^{12}$ $Hz.$


\begin{thebibliography}{99}
\bibitem{Nielsen-Chuang} M. A. Nielsen, and I. L. Chuang, \emph{Quantum
Computation and Quantum Information} (Cambridge University Press, Cambridge,
UK, 2000).

\bibitem{Bennett} C. H. Bennett, H. J. Bernstein, S. Popescu, and B.
Schumacher, Phys. Rev. A 53, 2046 (1996).

\bibitem{Peres} A. Peres, \emph{Quantum Theory: Concepts and Methods}
(Kluwer Academic Publishers, Boston, 1995).

\bibitem{Privman-Tolkunov-1} D. Tolkunov, and V. Privman, Phys. Rev. A 71,
060308(R) (2005).

\bibitem{Privman-Tolkunov-2} D. Tolkunov, and V. Privman, Proc. SPIE 5815,
187 (2005).

\bibitem{Bennett02} C. Bennett, D. DiVincenzo, J. Smolin, and W. K.
Wootters, Phys. Rev. A 54, 3824 (1996).

\bibitem{Storcz-Wilhelm} M. J. Storcz, and F. K. Wilhelm, Phys. Rev. A 67,
042319 (2003).

\bibitem{Ischi-Hilke-Dube} B. Ischi, M. Hilke, and M. Dube, Phys. Rev. B 71,
195325 (2005).

\bibitem{Yuetal01} T. Yu, and J. H. Eberly, Phys. Rev. B 66, 193306 (2002); 
\emph{ibid.} 68, 165322 (2003).

\bibitem{Yuetal02} T. Yu, and J. H. Eberly, Phys. Rev. lett. 93. 140404
(2004).

\bibitem{Leggett} A. J. Leggett, S. Chakravarty, A. T. Dorsey, M. P. A.
Fisher, and W. Zwerger, Rev. Mod. Phys. 59, 1 (1987); 67, 725(E) (1995).

\bibitem{split-operator01} S. K. Gray, and J. M. Verosky, J. Chem. Phys. 99,
8680 (1993).

\bibitem{split-operator02} S. Blanes, and P. C. Moan, Phys. Lett. A 265, 35
(2000).

\bibitem{Privman} L. Fedichkin, A. Fedorov, and V. Privman, Phys. Lett. A
328, 87 (2004); A. Fedorov, L. Fedichkin and V. Privman, J. Comp. Theor.
Nanosci. 1, 132 (2004); D. Tolkunov and V. Privman, Phys. Rev. A 69, 062309
(2004); V. Privman, J. Stat. Phys. 110, 957 (2003).

\bibitem{Concurrence} W. K. Wootters, Phys. Rev. Lett. 80, 2245 (1998); S.
Hill, and W. K. Wootters, Phys. Rev. Lett. 78, 5022 (1997).

\bibitem{chemphys} M. Goernale, M. Grifoni, and G. Sch\"{o}n, Chem. Phys.
268, 273 (2001).

\bibitem{chemphys295} Y. Makhlin, G. Sch\"{o}n, and A. Shnirman, Chem. Phys.
296, 315 (2003).

\bibitem{cpsun} C. P. Sun, S. Yi, and L. You, Phys. Rev. A 67, 063815 (2003).

\bibitem{Liang} X. T. Liang, and Y. J. Xiong, Physica B 362, 243 (2005).
\end{thebibliography}
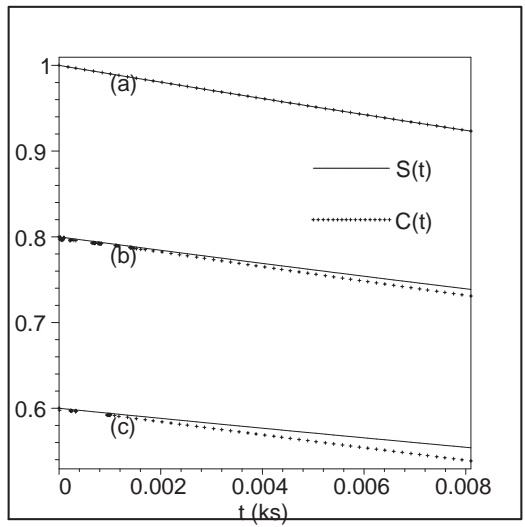
\end{document}